\documentclass[prb,nofootinbib,twocolumn,superscriptaddress]{revtex4} 

\usepackage{graphicx}
\usepackage{dcolumn}
\usepackage{bm}
\usepackage{threeparttable}
\usepackage{times}
\usepackage{mathptmx}
\usepackage{lscape}
\usepackage{natbib}
\usepackage{amsmath}
\usepackage{amssymb}
\usepackage{braket}

\def\degree{\kern-.2em\r{}\kern-.3em}

\begin{document}


\title{ First-principles Study on Formation of LPSO Structures for Ternary Alloys \\ Revisited from Short-range Order}

\author{Koretaka Yuge}
\affiliation{
Department of Materials Science and Engineering,  Kyoto University, Sakyo, Kyoto 606-8501, Japan\\
}%

\author{Hisashi Miyazono}
\affiliation{
Department of Materials Science and Engineering,  Kyoto University, Sakyo, Kyoto 606-8501, Japan\\
}%

\author{Ryohei Tanaka}
\affiliation{
Department of Materials Science and Engineering,  Kyoto University, Sakyo, Kyoto 606-8501, Japan\\
}%

\author{Tetsuya Taikei}
\affiliation{
Department of Materials Science and Engineering,  Kyoto University, Sakyo, Kyoto 606-8501, Japan\\
}%

\author{Kazuhito Takeuchi}
\affiliation{
Department of Materials Science and Engineering,  Kyoto University, Sakyo, Kyoto 606-8501, Japan\\
}%

\begin{abstract}
We investigate the formation of long-period stacking ordered (LPSO) structure for Mg-Y-Zn ternary alloys based on the short-range order (SRO) tendency of energetically competitive disordered phases. We find that unisotropic SRO tendencies for structures with stacking faults cannot be simply interpreted by arithmetic average of 
SRO for constituent fcc and hcp stackings, indicating that the SRO should be significantly affected by periodically-introduced stacking faults. We also find that SRO for neighboring Y-Zn pair, which should have positive sign to form specific L1$_2$ type cluster found in LPSO, is strongly affected by the distance between stacking faults: e.g., five atomic layer distance does not prefer in-plane Y-Zn pair, while seven atomic layer distance prefer both in- and inter-plane Y-Zn pair. These facts strongly indicate that ordering tendency for the Mg-Y-Zn alloy is significantly dominated by the stacking faults as well as their periodicity. We also systematically investigate correlation between SRO for other Mg-RE-Zn (RE = La, Tb, Dy, Ho, Er) alloys and the physical property of RE elements. We find that while SRO for RE-Zn pair does not show effective correlation with atomic radius, it has strong quadratic correlation with atomic radius considering unisotoropy along in- and inter-plane directions. 


\end{abstract}


\maketitle
\section{Introduction}
Long-period stacking ordered (LPSO) found in Mg-based ternary alloys, coexisting with $\alpha$-Mg phase, can be considered as promising candidate for 
light-weight structureal materials in next generation due to their outstanding tensile strength and ductility.\cite{lpso} 
Since LPSO exhibit modulation of structure (here, periodically introduced stacking faults) and that of concentration 
are synchronized, such peculiar microscopic structure is considered as key role for their specific structural properties: 
Especially, some of Mg-based alloys forming LPSO structures can have well-known L1$_2$ type clusters consisting of 8 RE atoms at vertex and rest six atoms at face-centered position of underlying fcc lattice at the stacking fault, where multiple clusters can exhibit specific in-plane ordering.\cite{l12} 
Following these stimulating facts, experimental\cite{okuda} as well as theoretical\cite{kimi1,kimi2} studies based on density functional theory (DFT) calculations have been amply perfomed to address the formation of LPSO structures mainly focuses on the ordering of specific L1$_2$ clusters. 
Particularly, DFT-based studies combined with empirical potentials have partiallye explained the cluster formation from solute atoms and in-plane ordering of the clusters themselves, while (i) how such ordering tendencies connects with perioedically-introduced stacking faults, and (ii) how the ordering tendencies depend on physical property for constituent elements that can/cannot form LPSO structures, are still under discussion.

Our recent DFT-based studies\cite{tana1,tana2} demonstrate the importance of considering the structural properties and energetics of disordered phases to address the stability of LPSO structure, where there found strong correlation between the SRO tendencies along nearest-neighbor (NN) coordination and whethere or not the ternary alloy can form LPSO structure, while the study did not address the above (i) and (ii). 
The present study addresses these points based on DFT calculations, finding the significant dependence of SRO on periodicity of stacking faults and physical quantity of constituent elements having strong correlation with the SRO tendencies. The details are discussed in the followings. 


\section{Methodology}
In order to quantitatively estimate temperature dependence of SRO including the effect of stacking faults for the present ternary alloy, we should in principle estimate energy for all possible atomic configuration, which is far from practical based on DFT calculation. Therefore, alternative approaches have been amply developed to overcome such problem, where generalized Ising model (GIM) is one of the most well-established and widely-used approach to accurately predict configurational energetics of alloys. 
However, in the present case, we should consider ternary atomic configuration on multiple stacking sequences (i.e., fcc, hcp and their mixed stacking) with various RE elements, which should required tremendous amount of computational costs. 
Very recently, we have developed theoretical approach,\cite{lsi,mers,rmx} enabling to quantitatively estimate SRO by using information only about two specially selected microscopic states, whose structure can be determined without any information about constituent elements, temperature or interactions. The present study employes the DFT energy for these special atomic configurations (SACS): One is called special quasirandom structure\cite{sqs} (SQS) corresponding to ideally random atomic configuration, and another is projection state (PS), including information about ordering tendency from random states with decresing temperature. Details of application of the theoretical approach is described in our previous papers. 

In the present study, to investigate the effect of periodicity in stacking faults on ordering tendency, we first consider the SRO on hcp stacking where stacking fault is introduced at every five or seven close-packed planes,  compared with arithmetic average of SRO for constituent fcc and hcp stackings. Since SACs depend only on their underlying lattice and selected coordination for SRO, we here numerically construct SACs along in- and inter-plane nearest-neighbor coordination for individual stacking sequences each of which composed of 80 atoms for five layers and 112 atoms for seven layers (i.e., 16 atoms at individual layer), with respective composition of HOGE and SOGE that takes as close as ideal composition of Mg-Y-Zn alloys can form LPSO structure of 18R type. 
In order to obtain total energy of SACS on multiple stacking sequences with various RE elements as described above, first-principles calculations based on DFT, the Vienna Ab-initio Simulation Package (VASP) code,\cite{vasp1,vasp2} is performed. All-electron Kohn-Sham equations were solved based on the projector augmented-wave (PAW) method\cite{paw1,paw2} with the generalized-gradient approximation of the Perdew-Burke-Ernzerhof\cite{pbe} (GGA-PBE) form to the exchange-correlation functional. We set the plane-wave cutoff energy of 350 eV throughout the calculations, and Brillouin zone sampling was performed using the Monkhorst-Pack scheme.\cite{mp} The k-point sampling is on the $4\times 4\times 4$ and $4\times 4\times 2$ for 80- and 112-atom cell. 

Atomic radius for RE, $R$, is estimated in two ways. One is that we consider atomic radius of RE isotropic. Under this condition, atomic radius is estimated by comparing the volume of 80 Mg atom cell where one Mg is replace by Re with that of pure 80 Mg atom cell: We apply the Vegard's law to determine $R$ using the differences in volume of the two cells.  Another way is that we consider atomic radius of RE unisotropic along in- and inter-plane coordination, $R^{\parallel}$ and $R^{\perp}$: In a similar fashion to the isotropic case, we can determine $R^{\parallel}$ and $R^{\perp}$ using the differences in in-plane area and lattice constant along inter-plane coordination under the condition of Vegard's law. 


\begin{figure}[h]
\begin{center}
\includegraphics[width=0.83\linewidth]
{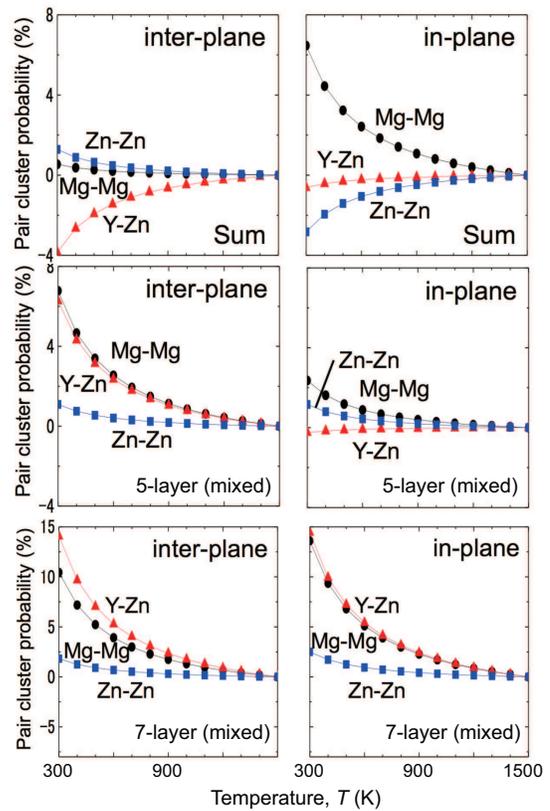}
\caption{SRO tendency in terms of probability for Mg-Mg, Zn-Zn and Y-Zn neighboring pairs along inter- and in-plane directions as a function of temperature, measured from that at $T=1500$ K. Upper two figures: SRO for arithmetic average of fcc and hcp stackings. Middle two: SRO for 5-layer mixed (fcc and hcp) stacking. Lower two: SRO for 7-layer mixed staking. }
\label{fig:sro}
\end{center}
\end{figure}

\begin{figure}[h]
\begin{center}
\includegraphics[width=0.9\linewidth]
{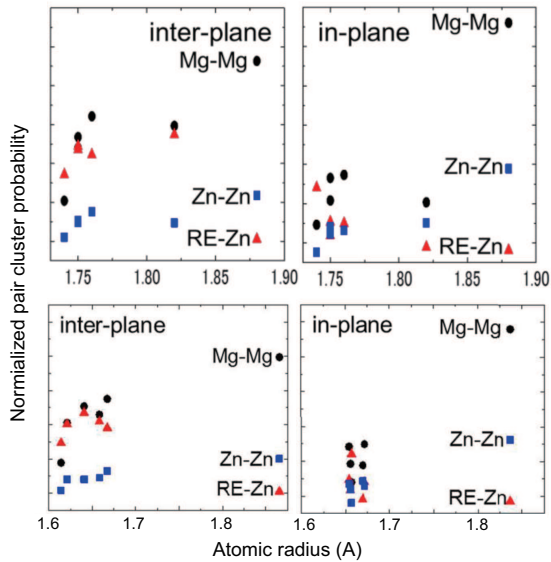}
\caption{SRO tendency for neighborin in- and inter-plane directions for Mg-RE-Zn alloys (RE = Y, La, Tb, Dy, Ho, Er) in terms of isotropic (upper figures) and unisotropic (lower figures) atomic radius obtained from DFT calculations. }
\label{fig:r}
\end{center}
\end{figure}

\section{Results and Discussions}
Figure~\ref{fig:sro} shows the predicted SRO tendency in terms of probability for Mg-Mg, Zn-Zn and Y-Zn neighboring pairs along inter- and in-plane directions as a function of temperature, measured from that at $T=1500$ K. Hereinafter we mainly focus on SRO for Y-Zn pairs. When we see the upper two figures for arithmetic average, SRO for Y-Zn pair exhibits negative sign along both in- and inter-plane directions, whose magnitude is enhanced with decrease of temperature. This certainly indicates that when stacking faults are infinitely separated (i.e., no interaction between stacking faults), forming Y-Zn neighboring pair is thermodynamically disfavored, which is in contrast to the specific L1$_2$ cluster found in LPSO having positive sign of SRO for Y-Zn along both in- and inter-plane directions. 
When we introduce periodic stacking faults at each 5 layer (middle two figures), SRO for Y-Zn pair along inter-plane direction significantly changes: the SRO exhibits positive sign, which is enhanced with decrease of temperature. This indicates that when stacking faults are periodically introduced, Y-Zn pair along inter-plane direction is thermodynamically preferred. 
Meanwhile, when we see in-plane SRO, this tendency does not hold true: Even when the stacking fault is introduced, SRO for Y-Zn pair does not show any significant change, still remaining 
negative sign. To form L1$_2$ cluster in LPSO, in addition to Y-Zn pair, SRO for Mg-Mg and Zn-Zn pair along both in- and inter-plane direction have positive sign: With this consideration, for stacking faults at every 5-layer, all neighboring pairs other than in-plane Y-Zn pair satisfy this condition. 
With the results of upper and middle two figures in Fig.~\ref{fig:sro}, when distance between periodic stacking faults increases, simple expectation is that inter-plane Y-Zn pair becomes thermodynamically disofavored, and in-plane Y-Zn remains disfavored. However, when we practically introduce periodic stacking faults at every 7 layer (lower two figures), the predicted SRO 
exhibit completely opposite behavior: Preference of inter-plane Y-Zn pair is more enhanced at 7-layer stacking faults than that at 5-layer, and SRO for in-plane Y-Zn exhibit positive sign whose magnitude is significantly enhanced with decrease of temperature. Furthermore, SRO for all neighboring pairs exhibit positive sign, which has the same tendency of SRO for the LPSO structure. 
These results certainly indicates that there should be "optimal" distance between stacking faults to exhibit ordering tendency precursor to form L1$_2$ cluster in LPSO for Mg-Y-Zn ternary alloys.  

\begin{figure}[h]
\begin{center}
\includegraphics[width=0.6\linewidth]
{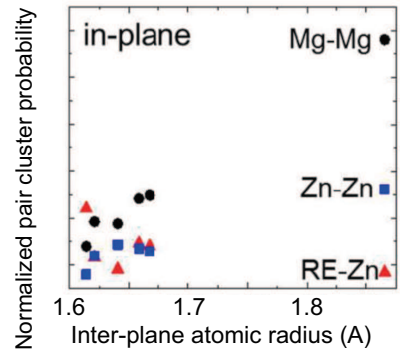}
\caption{In-plane SRO tendency for Mg-RE-Zn alloys in terms of inter-plane atomic radius of RE. }
\label{fig:sroin}
\end{center}
\end{figure}

Next, in order to systematically see how the in- and inter-plane SRO (especially for RE-Zn pair) correlates with physical property for constituent RE element, we show in Fig.~\ref{fig:r}  
in- and inter-plane SRO for RE-Zn pair of stacking fault at every 5 layer whose unisotropy for Mg-Y-Zn alloy exhibit highest, in terms of isotropic (upper figures) and of unisotropic (lower figures) atomic radius of RE. 
From upper figures, we can see that RE = La, that do not form LPSO, has significantly larger atomic radius than other RE elements forming LPSO structures.
Other than La, we can see no significant linear or quadratic correlation between SRO and isotropic atomic radius along in- or inter-plane direction.  
This can be a natural result since from Fig.~\ref{fig:sro}, SRO for RE-Zn pair exhibit strong unisotropy (its sign depends on the direction), while typical atomic radius does not explicitly contain such unisotropic information. 
When we decompose atomic volume into individual unisotroipic atomic radius, correlations for inter-plane SRO become much clearer: inter-plane SRO for RE-Zn pair exhibits convex upward with respect to inter-plane atomic radius of RE other than La, while in-plane SRO still does not show effective correlation. The latter can be naturally accepted since differences in in-plane atomic radius other than La is significantly small compared with that for inter-plane atomic radius, resulting in less correlation with the SRO. 
When we interpret that inter-plane atomic radius reflects the effects of interactions between stacking faults, the effects are expected to change the in-plane SRO tendency. 
Under this consideration, we show in Fig.~\ref{fig:sroin} in-plane SRO tendency in terms of inter-plane atomic radius of RE. We can see quadratic correlation with converx downward between 
atomic radius and SRO: This is the opposite SRO tendency along inter-plane directions with convex upward correlation. 
These facts strongly indicate that unisotropy in atomic radius, especially along inter-plane direction, has significant connection with in- and inter-plane SRO tendency for RE-Zn pair and would reflects the effects between periodic stacking faults, which should be additional fundamental key to systematically understand the formation of LPSO structure for Mg-based ternary alloys. 
Quantitative relationship between unisotropy of atomic radius and interactions between stacking faults should be further investigated in our future study. 

\section{Conclusions}
Formation of LPSO structure for Mg-based ternary alloys is investigated based on SRO tendency of neighboring atomic pairs, where effects of introducing periodic stacking faults as well as correlation with unisotropy in atomic radius for RE elements are considered. We find that SRO on hcp lattice with periodic stacking faults cannot be simply described by arithmetic average of SRO on constituent hcp and fcc lattice, indicating profound effects of stacking faults on SRO for the present alloy: The SRO tendency, especially the preference of RE-Zn pair, can be reversed when distance between stacking faults changes. We also find that SRO of RE-Zn pair along in- and inter-plane direction have strongly quadratic correlation with inter-plane atomic radius of RE, indicating that relationship between unisotropy of atomic radius and interactions between stacking faults should be further investigated to clarify the key role to forming specific in-plane ordering found in LPSO structures.

 \section*{Acknowledgement}
This work was supported by a Grant-in-Aid for Scientific Research (16K06704) from the MEXT of Japan, Research Grant from Hitachi Metals$\cdot$Materials Science Foundation, and Advanced Low Carbon Technology Research and Development Program of the Japan Science and Technology Agency (JST).

\end{document}